# Synergically enhanced viscoelastic behavior of binary nanocarbon based polyurethane hybrid nanocomposite foams

Amir Navidfar [a,b*], Osman Bulut [c], Tugba Baytak [c], Hikmet Iskender [d], Levent Trabzon [a,b]

[a] *Faculty of Mechanical Engineering, Istanbul Technical University, Istanbul, Turkey;*

[b] *Nanotechnology Research and Application Center, Istanbul Technical University, Istanbul, Turkey;*

[c] *Faculty of Civil Engineering, Istanbul Technical University, Istanbul, Turkey;*

[d] *Disaster Management Institute, Istanbul Technical University, Istanbul, Turkey;*

## Abstract

Carbon-nanofillers are known for improving the desired properties of polymers. The dispersion quality of nanofillers in the matrix is vital for the fabrication of high-performance nanocomposites. An effective approach for improving dispersion states of multi-walled carbon nanotubes (MWCNT) and graphene nanoplatelets (GNPs) was employed via hybrid inclusion of the nanofillers in polyurethane matrix to further enhancing viscoelastic properties. Nanocomposites based on MWCNTs, two groups of graphene and hybrid MWCNT/graphene with varied weight fractions and ratios were fabricated via a simple, quick and scalable approach. Dynamic mechanical analysis results indicated an improvement of up to 86% in storage modulus at 25ºC for hybrid MWCNT/GNP-S750 at only 0.25 wt% loading, whereas solely MWCNTs and graphene nanocomposites showed 9% and 15% enhancement at the same content, respectively. The glass transition temperature value was enhanced by about 9.5 ˚C with 0.25 wt% inclusion of well-dispersed three-dimensional MWCNT/GNP-S750 structure, which disclosed a noticeable synergistic effect in thermomechanical properties.

[*]**Correspondence to: Amir Navidfar (Email: Navidfar@itu.edu.tr)**

# 1. Introduction

Carbon nanofillers such as multi-walled carbon nanotubes (MWCNTs) and graphene nanoplatelets (GNPs) are one of the most promising nanomaterials. Their extraordinary mechanical strength, thermal and electrical properties make them highly demanding in reinforcing polymeric materials [1-6]. However, agglomerations and the restacking tendency of these nanofillers enhanced dispersion difficulty in the polymer matrix [7-10]. Hybridization of the one-dimensional (1D) MWCNTs and two-dimensional (2D) graphene with different geometry can overcome this dispersion challenge. The synergistic effect between CNTs and graphene has been shown to enhance mechanical properties [11-13], thermal [14, 15] and electrical conductivity [16-18]. Bagotia et al. [12] concluded an enhancement in mechanical properties, electrical conductivity and EMI shielding of nanocomposites with graphene/MWCNT hybrid fillers. In the hybrid graphene/CNT, it is believed that 1D CNTs bridge the adjacent graphene to form a three-dimensional (3D) network architecture. In this condition, graphene and CNTs exhibit a stronger synergistic effect in increasing the contact area between nanofillers and polymer matrix [19, 20], which formed an effective network to strain transferring [21].

The energy efficiency requirements increasingly demand lightweight structures such as polyurethane (PU) foams. The simple fabrication of PU structures and composites in a single processing step is the main advantage of PU technology [22]. Due to their low density, PU foams have been used in versatile applications such as insulation, packaging, seating and sound-absorbing materials [23, 24]. However, PU foams have some disadvantages, such as low mechanical properties and deficient structural stability in high temperatures, which hinders their widespread application in the automobile and aerospace industries [25]. Hence, graphene [26, 27] and MWCNTs [7, 21, 28, 29] have been employed in the development of mechanically enhanced and thermally stable PU nanocomposites[30]. In our previous work, the tensile strength of PU enhanced synergically up to 43% with the addition of graphene/MWCNT hybrids, while the strength of PU/MWCNTs and PU/graphene improved 19% and 17% at the same content, respectively [13]. According to the numerous applications of PU, dynamic mechanical analysis (DMA) as a tool in the evaluation of nanocomposite structures is critical. DMA provides the mechanical properties of viscoelastic materials under periodic loading and thermal changes for a diversity of specimen geometries. It is used to determine a variety of major material parameters containing storage ($E'$) and loss modulus ($E''$), damping factor (tan δ), glass transition temperatures ($T_g$), creep compliance and relaxation modulus [31]. It has been verified to be an effective technique to investigate the relaxations in polymers and thereby materials behavior under several conditions of temperature, stress, nanocomposites structure and their influence in evaluating the mechanical properties. Numerous works have been investigated storage modulus, loss modulus and tan δ at a temperature range for PU nanocomposites [23, 29, 32]. However, there are limited studies on the viscoelastic properties of MWCNTs/graphene hybrid nanocomposites with a synergistic effect. Pokharel et al. [32] studied the DMA behavior of graphene, graphene oxide (GO) and functionalized graphene sheet (FGS) reinforced PU nanocomposites at 2 wt% concentration. Their results showed superior thermomechanical properties of FGS/PU than other nanofillers incorporation in PU, due to improved interface in FGS/PU nanocomposite. Roy et al. [29] fabricated thermoplastic PU via solution mixing method with chemically reduced GO and MWCNT hybrids, which the storage modulus at the glassy region enhanced by 51% at 0.25 wt% loadings. Wang et al [33] indicated an outstanding dynamic mechanical properties enhancement in waterborne epoxy via introducing graphene oxide/carbon nanotube hybrid nanofillers.

Nevertheless, their works were focused on single nanofillers inclusion or limited study of hybrid nanofillers on bulk polymeric matrices instead of cellular PU foams.

The present study focused on the viscoelastic behavior of PU foam nanocomposites reinforced with graphene/MWCNTs hybrid nanofillers. Commercially available MWCNTs and GNPs with different dimensions and specific surface areas were used to investigate the synergistic effect of carbon nanofillers on the thermomechanical properties of PU. Varied nanofiller ratios and concentrations were studied to find the highest synergy, in which 86% improvement in storage modulus is achieved at 0.25 wt%, relative to pure PU. This work also reveals the capability to reach these significant results in very low loadings of low cost commercially available GNPs and CNTs. Successful dispersing of these hybrid carbon nanofillers in the polymeric matrix can pave the way toward low cost and efficient composite fabrication.

## 2. Materials and Experimental Study

### 2.1 Materials

For the fabrication of polyurethane foams, the polyol and isocyanate with a weight ratio of 1:1.25, as suggested by the manufacturer were mixed. The density and viscosity of the polyol (at 25°C) are 1.11 g/cm$^3$ and 600 ± 200 MPa.s, respectively. The hydroxyl number of the polyol is 300 mg KOH/g. The density and viscosity of isocyanate (at 25°C) are 1.23 g/cm$^3$ and 210 MPa.s, respectively. The NCO content of isocyanate is about %30.8 - %32. Carbon nanofillers were purchased from Nanografi Co.Ltd, which the properties and geometries of used MWCNTs and graphene are tabulated in Table 1. MWCNTs (purity ≥ 92%) were grown by chemical vapor deposition.

**Table 1**

Properties and geometry of nanofillers.

| Nanofillers | Diameter (nm) | Thickness/Length (nm) | Specific Surface Area (m$^2$/g) |
|---|---|---|---|
| **MWCNT** | 8-10 | L =1000-3000 | 290 |
| **GNP-L150** | 24000 (Large) | t = 6 | 150 |
| **GNP-S750** | 1500 (Small) | t = 3 | 750 |

## 2.2   Fabrication of Nanocomposites

MWCNTs were functionalized with hydrogen peroxide (H$_2$O$_2$) before the fabrication of PU foams [13, 15, 21]. Different contents (0.25, 0.50 and 0.75 wt%) and ratios (3:1, 1:1 and 1:3) of MWCNTs and graphene were first mixed with polyol at 200-2000 rpm for 5 min using overhead stirrer equipment. Polyol and carbon nanofiller combinations were ultrasonically dispersed for 5 min using an ultrasonic bath and then stirred at 2000 rpm for 5 min. The isocyanate was finally added to the mixture, which was stirred for 20 s and then expanded in a two-part wooden mold to form free-rise PU foam. Cured nanocomposites were cut with a thickness of 10 mm using a lathe machine.

## 2.3   Characterization

DMA technique was used to measure the viscoelastic properties of fabricated nanocomposites using TA Instruments Q800 DMA at a loading frequency of 1 Hz from 25 °C to 180 °C with a heating rate of 3 °C/min in single cantilever mode. A desktop CNC machine was used to prepare desired DMA specimens with a size of 25 mm x 15 mm x 2 mm. The glass transition temperatures ($T_g$) of the samples were also obtained by DMA tests. To examine the repeatability of the DMA results, three specimens were prepared and tested for each type of nanocomposite.

Thermogravimetric analysis (TGA) was carried out under a nitrogen atmosphere on a Q600, TA Instruments with a heating rate of 5 °C/min. Transmission electron microscopy (TEM) images of carbon nanofillers and their hybrids were taken using a Hitachi HighTech HT7700 with an acceleration voltage of 120 keV. Besides, scanning electron microscopy (SEM) experiments of fabricated PU nanocomposites were done by an FEI - NOVA NanoSEM 450.

## 3. Results and Discussions

### 3.1 Material Characterization of Carbon Nanofillers and Their Nanocomposites

Several techniques such as Raman Spectroscopy, TGA, SEM and TEM were utilized for the characterization of MWCNTs, graphene and their nanocomposites. The decomposition behavior of fabricated nanocomposites was investigated using thermogravimetric analysis. TGA and derivative thermogravimetric (DTG) curves, as well as degradation temperature at 50% weight losses and charred residue of the nanocomposites, are illustrated in Fig. 1(a-d). All nanocomposites at 0.25 wt% except PU/MWCNT exhibited higher thermal stability, relative to neat PU. The decomposition temperature of hybrid MWCNT/GNP-S750 nanocomposite enhanced due to nanofillers synergistic effect. However, the enhancement in decomposition temperature was higher for the nanocomposite with solely GNP-S750, compared to hybrid MWCNT/GNP-S750 (Fig. 1(a) inset), which is attributed to the higher thermal conductivity of hybrid MWCNT/GNP-S750 in the PU matrix, in comparison with PU/GNP-S750. Hybrid MWCNT/GNP-S750 nanocomposite lost weight at lower temperatures compared to PU/GNP-S750, owing to higher heat flow throughout the specimen. This means that thermal conductivity enhancement dominated improved nanofillers dispersion in hybrid nanocomposites [34]. Fig. 1(c) exhibited that the thermal stability of PU improved by 3 ˚C, 7.3 ˚C and 14.5 ˚C with MWCNT/GNP-S750, GNP-L150 and GNP-S750 nanofillers at 50% weight loss, respectively. It is also noted that charred residue of PU at 600 ˚C improved, especially in GNP-S750 included nanocomposites, as shown in Fig. 1(d). This may be due to the homogeneous distribution of GNP-S750, as well as their strong interaction with the polymer matrix caused by their higher specific surface area [29, 35, 36]. According to Fig. 1(d), among all nanocomposites, PU/GNP-S750 showed the highest residue of 24.4 wt% at 600 ˚C. It is reported that graphene revealed a notable barrier effect to delay the thermal degradation of nanocomposites by means of hindering the exit of volatile products during the degradation [37, 38]. The derivative thermograms (Fig. 1(b)) indicates that the maximum decomposition rate is higher for PU/MWCNT. However, the peak intensity decline in other nanocomposites, especially in GNP-S750 and MWCNTs+GNP-S750 based nanocomposites. Raman spectra are commonly used to characterize quality, crystal structure, disorder and defects of carbon-based materials such as CNT and graphene quickly and non-destructively. Fig. 1(e) exhibited D-bands, G-bands and 2D-bands of the carbon nanofillers, which are relevant to their defects and carbon networks [39]. The $I_D/I_G$ ratio of GNP-S750 ($I_D/I_G = 0.49$) indicated a higher value than those of GNP-L150 ($I_D/I_G = 0.08$), demonstrating extra defects and porous graphene. The lower flakes dimension and more functional groups on the surface and edge of GNP-S750 enhanced the disorder degree [36]. Also, the D-band to G-band intensity ratio of MWCNTs ($I_D/I_G = 1.19$) was higher than those of graphene, which was ascribed to $sp^3$ defects and twist network of carbon nanotubes [40].

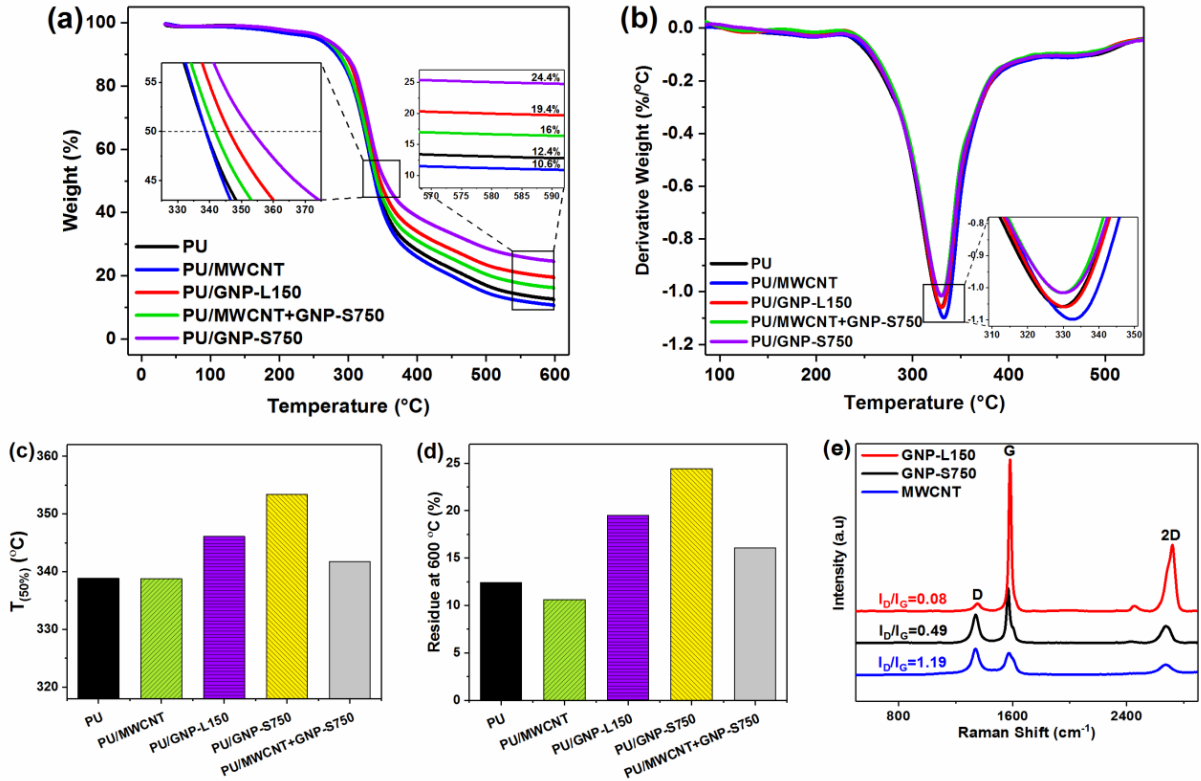

Fig. 1. (a) TGA and (b) DTG curves, (c) temperature at 50% weight loss, (d) residue at 600 °C of nanocomposites and (e) Raman spectra of carbon nanofillers.

## 3.2 Morphological Characterization

The TEM micrographs of the graphene and MWCNTs (Fig. 2(a-c)) displayed wrinkled and crumpled structure of graphene nanosheets, as well as randomly organized and aggregated MWCNTs due to inter-molecular Van der Waals interaction, which were entangled. Fig. 2(d-g) exhibits SEM images of PU with various carbon nanofillers, as well as GNP-S750/MWCNTs hybrids at 0.25 wt%. Several agglomerations are apparent in PU/MWCNTs, while a well-dispersed hybrid MWCNTs/GNP-S750 nanocomposite is obvious, according to Fig. 2(g). 1D MWCNTs connected to the 2D graphene surface and formed a 3D architecture, which hindered agglomerations [41]. This 3D structure expands the contact surface areas among MWCNTs/GNP-S750 nanofillers and the PU matrix that is favorable in stress transferring.

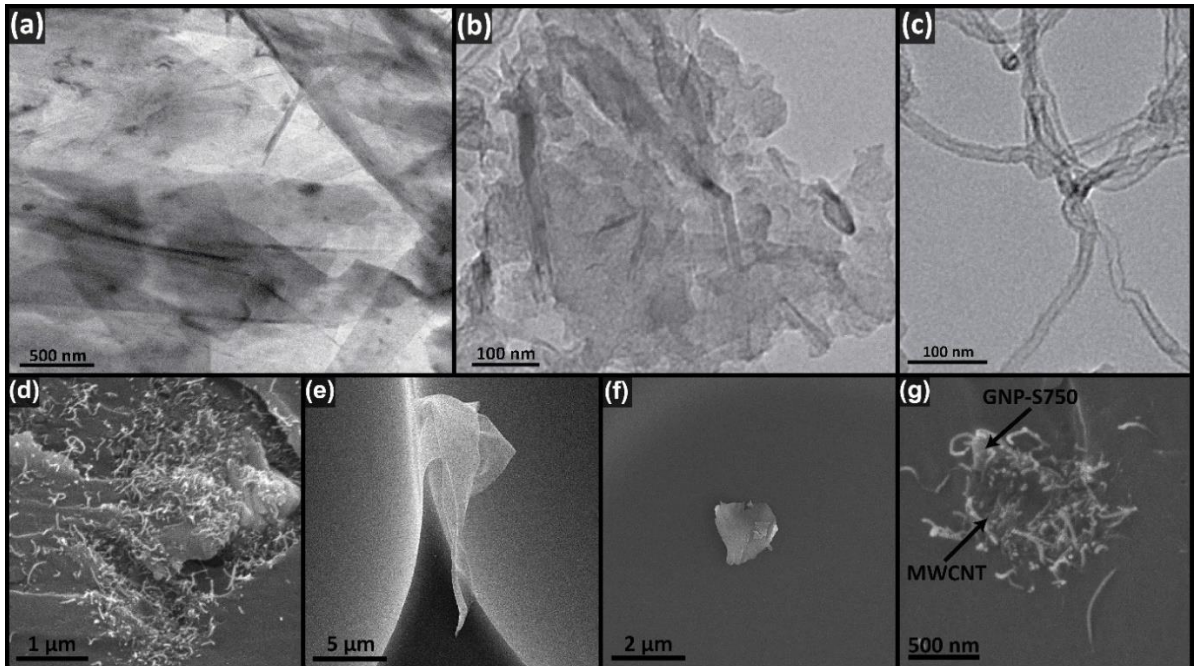

Fig. 2. TEM images of (a) GNP-L150, (b) GNP-S750, (c) MWCNTs and SEM of (d) PU/MWCNTs, (e) PU/GNP-L150, (f) PU/GNP-S750 and (g) PU/MWCNT+GNP-S750 nanocomposites.

### 3.3 Viscoelastic Properties

#### 3.3.1 Storage Modulus

DMA measurements reveal the viscoelastic properties of the nanocomposites. Fig. 3 exhibits the variation of dynamic storage modulus ($E'$) of PU with the change of temperature and its nanocomposites including carbon nanofillers, which implies nanocomposites stiffness. The results showed that the storage modulus declined very slowly in the temperature range of 25–100 °C but after 100 °C the storage modulus diminished abruptly due to the transition from glassy to rubbery state [23]. The storage modulus improved in the glassy region (about 25 ˚C) with the carbon nanofillers loadings, being higher for the nanocomposites at 0.75 wt% contents owing to the restriction in mobility of polymer chains provided by the nanofillers (Fig. 3(d)). The resilient properties of MWCNTs and graphene improved the storage modulus by transferring the stress under loading, besides acting as additional physical crosslinks [42]. The storage modulus of PU was improved up to around 16%, 19%, and 22% for MWCNTs, GNP-L150, and GNP-S750 nanocomposites at 0.75 wt %, respectively. Higher modulus improvement in GNP-S750 based nanocomposites is ascribed to the homogeneous distribution of GNP-S750, as well as the strong interaction with the polymer matrix due to their higher specific surface area and lower flake size [29, 36]. Moreover, GNP-S750 with smaller thickness and diameter could efficiently enter among polymer chains and restrict their mobility, as compared to GNP-L150 [36].

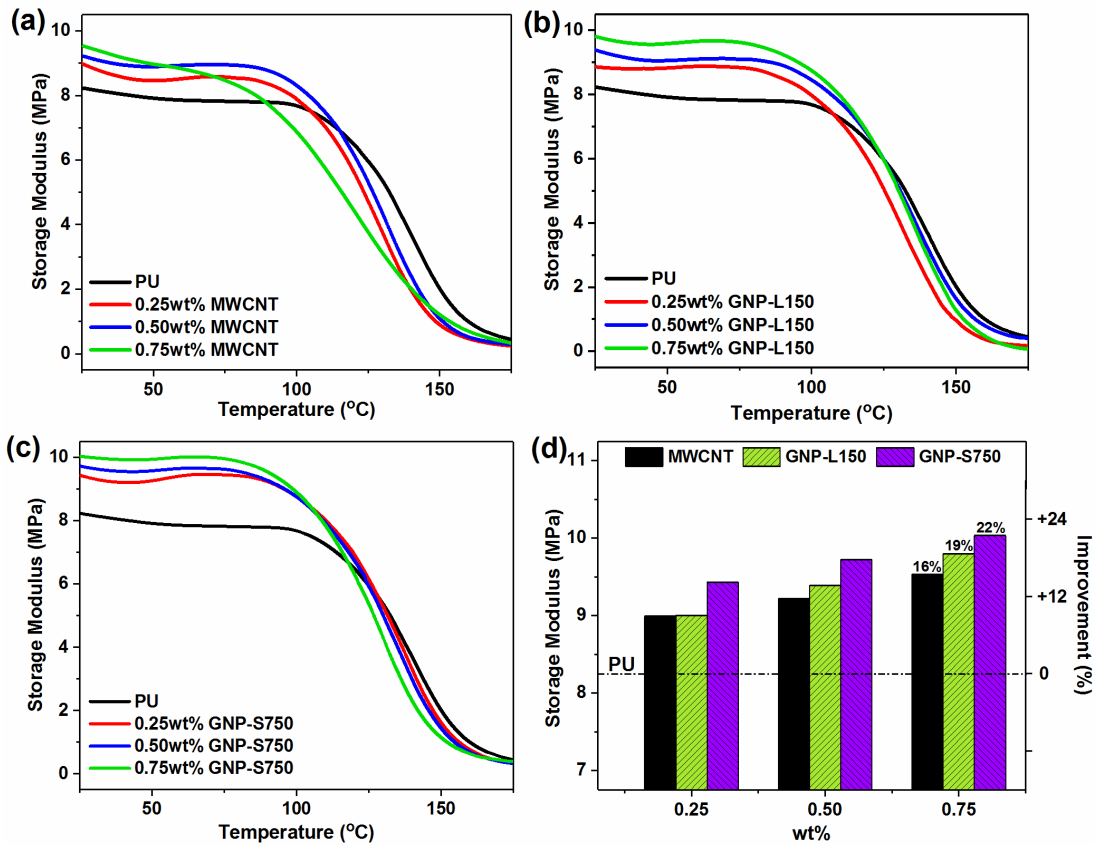

Fig. 3. DMA storage modulus versus temperature of (a) MWCNTs, (b) GNP-L150 and (c) GNP-S750 nanocomposites and (d) storage modulus in the glassy region (25°C).

Fig. 4(a-d) illustrates DMA storage modulus versus temperature of hybrid nanocomposites in comparison with single MWCNTs and single graphene based PU. Besides, the storage modulus in the glassy region (25°C) is summarized in Fig. 4(e-f). The results revealed that hybrid nanofillers inclusion in PU promoted the storage modulus as a result of their better dispersion into the polymer

matrix. It is evident that storage modulus of PU with hybrid MWCNT/GNP-S750 (1:1) dramatically enhanced up to about 86% at 0.25 wt% in the glassy region (25˚C), relative to pure PU (Fig. 4(a) and (e)), while the moderate improvement of 22% can be seen in MWCNTs/GNP-L150 (1:3) based hybrid nanocomposites at the same content. According to Fig. 4(f), maximum improvement of storage modulus using MWCNTs and GNP-L150 hybrids can be obtained about 29% at 0.75 wt%. As a result, MWCNTs with GNP-S750 showed a remarkable synergistic effect in enhancing the storage modulus of PU at a low nanofillers content of 0.25 wt%. Two-dimensional GNP-S750 with lower flake dimension and higher SSA and defects has more ability to be interconnected by 1D MWCNTs to develop a 3D well-dispersed nanofillers network [34]. This 3D hybrid architecture promoted the interaction of nanofillers with PU matrix, in which the enhanced contact area facilitated the load transferring through the nanofillers [13, 15, 29].

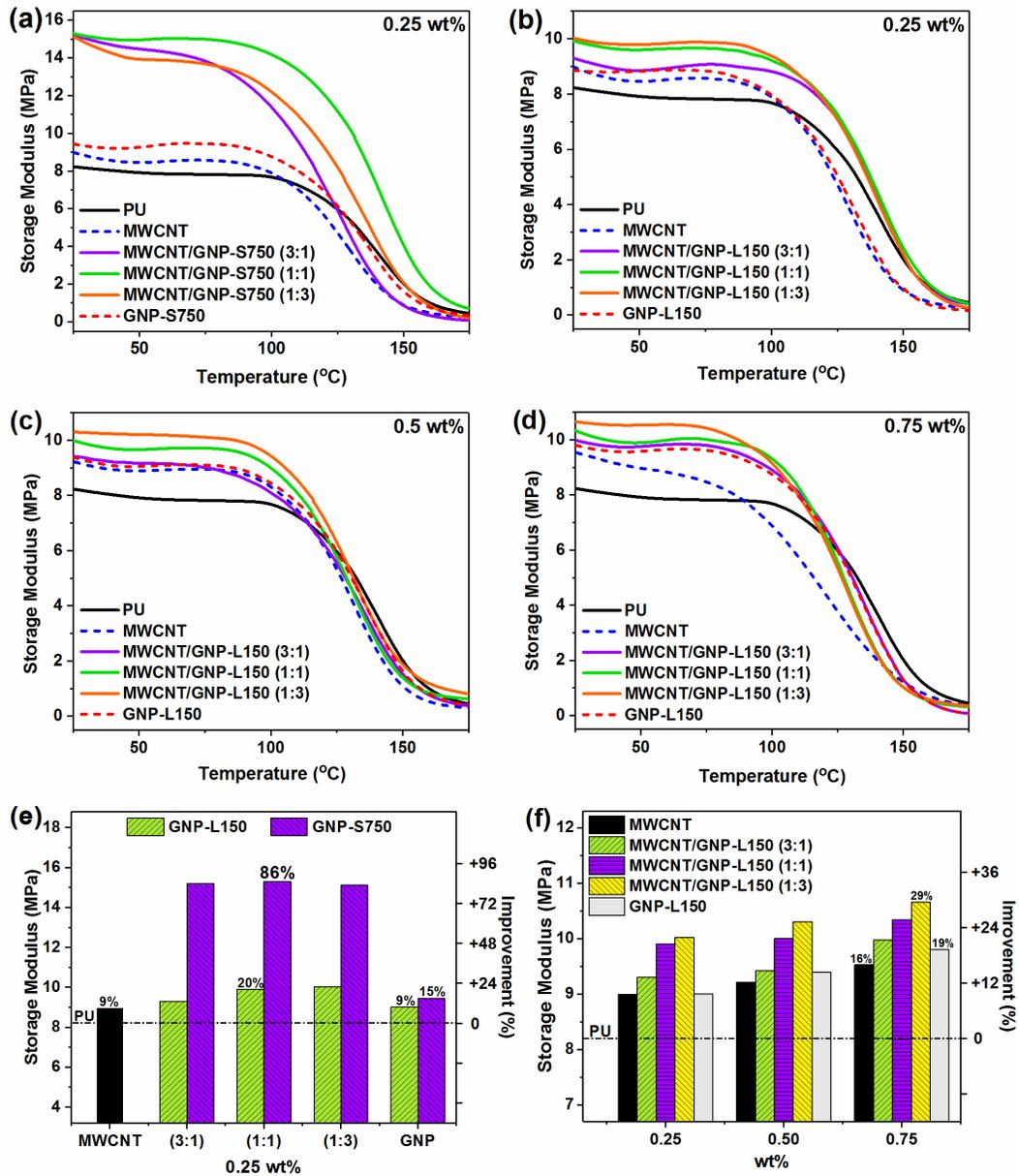

Fig. 4. DMA storage modulus of (a-d) hybrid MWCNTs/GNPs nanocomposites versus temperature and (e-f) storage modulus in the glassy region (25°C).

Excitingly, the TEM image of the hybrid MWCNTs/GNP-S750 in Fig. 5(d) demonstrates the uniform dispersion of carbon nanotubes on the GNP-S750 surface to connect the adjacent graphenes that made a 3D interconnecting structure. Hence, the hybrid MWCNT/GNP-S750 nanofillers are anticipated to provide an outstanding improvement in the mechanical and thermal properties of nanocomposites [43, 44]. To better understand the mechanism of hybrid nanocomposites in various strain conditions in DMA, a schematic illustration about the change of the 3D interconnecting network of GNP-S750/MWCNTs hybrids in PU matrix before loading, under loading and in relaxation time is displayed in Fig. 5(a-c). In the primary state, well-dispersed nanofillers were formed by the interconnected MWCNTs and graphene. When the graphene /CNTs nanocomposites were stretched to a strain, the distance among nanofillers enhances (red lines), causing the higher modulus. However, the initial nanofillers condition could be found again after the relaxation. Consequently, this 3D interconnected structure caused to enhance the storage modulus of the hybrid nanocomposites under periodic strain.

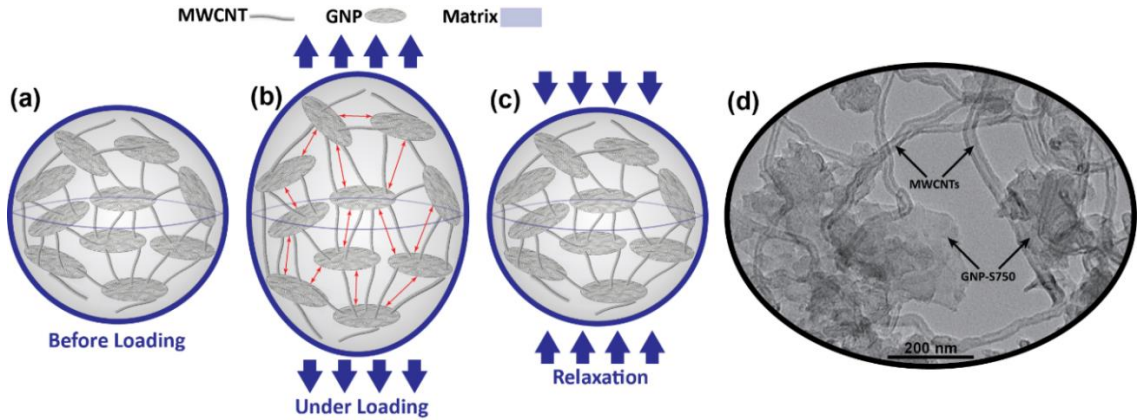

Fig. 5. (a-c) Schematic of the change of graphene/MWCNTs hybrid network under different strains and (d) TEM image of hybrid MWCNTs/GNP-S750.

### 3.3.2 Tan δ and Glass Transition Temperature ($T_g$)

The temperature-dependent tan δ (damping factor) curves of pristine PU and carbon nanofiller based nanocomposites are displayed in Fig. 6(a-c), which represents the balance of dissipated energy (viscous behavior) and stored energy (elastic behavior). The temperature value in the peak of tan δ curves suggested the glass transition temperatures ($T_g$), as summarized in Fig. 6(d). PU/GNP-S750 nanocomposite at 0.5 wt% showed the highest $T_g$ value (155.5 °C) among all nanofillers with 5.8 °C enhancement compared to pure PU (149.7 °C) due to the better interfacial adhesion of the nanofillers and the polymer matrix [45]. This $T_g$ enhancement can be attributed to the smaller diameter of GNP-S750, which reduced the polymer chains mobility through efficiently entering to polymer chains, as well as better dispersion in PU matrix due to their higher specific surface area [36]. PU with GNP-L150 at 0.25 wt% indicated a 2.1 °C shift in $T_g$ value, in comparison with PU, but the amount of $T_g$ declined in higher GNP-L150 loadings owing to their poor dispersion state in higher concentrations [7, 13, 21]. The results also presented reduced $T_g$ with MWCNTs loadings, which is caused by large agglomerations as illustrated in SEM images (Fig. 2(d)). These agglomerates decreased chain interaction, declined the crosslink density and enhanced free volume in the nanocomposites [36]. As a result, graphene is more efficient than MWCNTs in enhancing the $T_g$ value of PU [46].

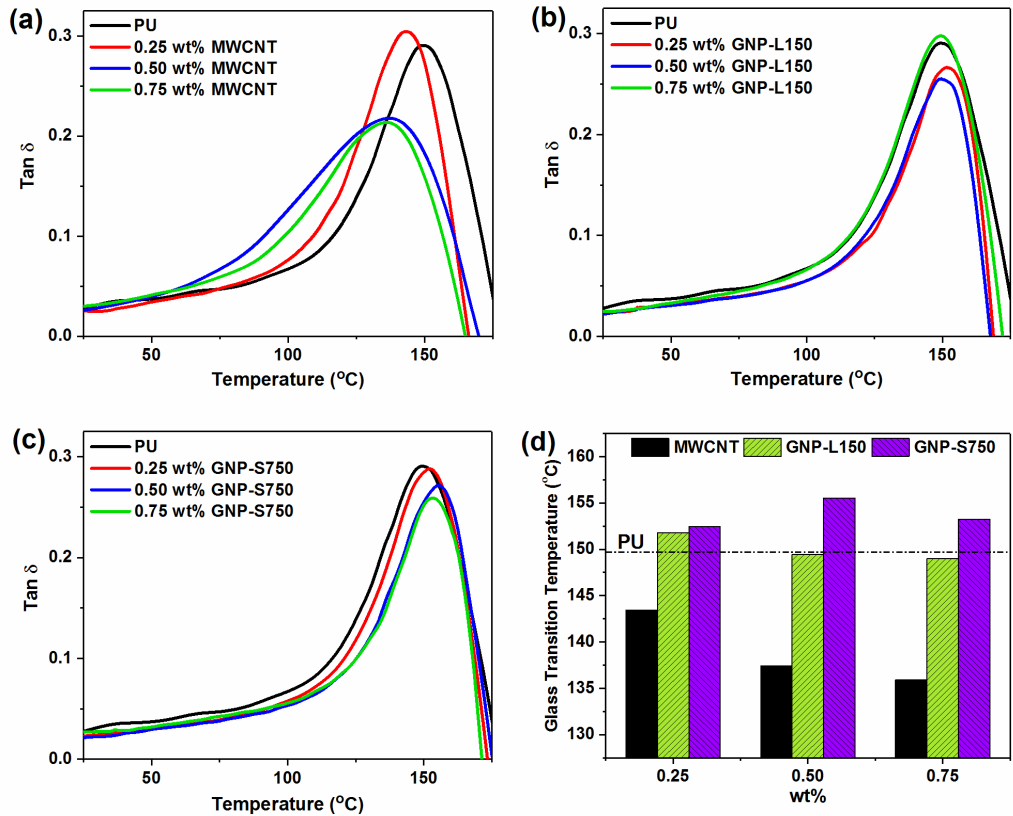

Fig. 6. Tan δ versus temperature of (a) MWCNTs, (b) GNP-L150 and (c) GNP-S750 nanocomposites and (d) glass transition temperature of nanocomposites.

Fig. 7(a-d) illustrates tan δ versus temperature curves of hybrid nanocomposites, in comparison with the single nanofillers included nanocomposites and neat PU, which provides the information regarding the ratio of viscous to elastic properties of nanocomposites. The glass transition temperature of pure PU measured $T_g$ = 149.7 ˚C. The results revealed that $T_g$ values are shifted to higher temperatures in hybrid nanocomposites, which surpassed the $T_g$ of single nanofillers loaded nanocomposites [47]. The nanocomposites with MWCNTs/GNP-L150 hybrids showed a $T_g$ shift of 5.6 ˚C at 0.25 wt% with an MWCNT: graphene ratio of 1:3. MWCNT/GNP-S750 (1:1) hybrid nanocomposite synergistically exhibited the highest $T_g$ value with 9.5 ˚C enhancement (159.2 ˚C), relative to $T_g$ of pure PU (Fig. 7(e)). This temperature shift caused by the immobilization effect of hybrid nanofillers on the polymer chains [38]. Homogeneous dispersion was contributed to a higher $T_g$ because of increased contact area with polymer chains to constrain their mobility [46]. Thus, the 3D well-dispersed hybrid MWCNT/GNP-S750 structure results in a rise of the surface area with the PU and a higher $T_g$ value.

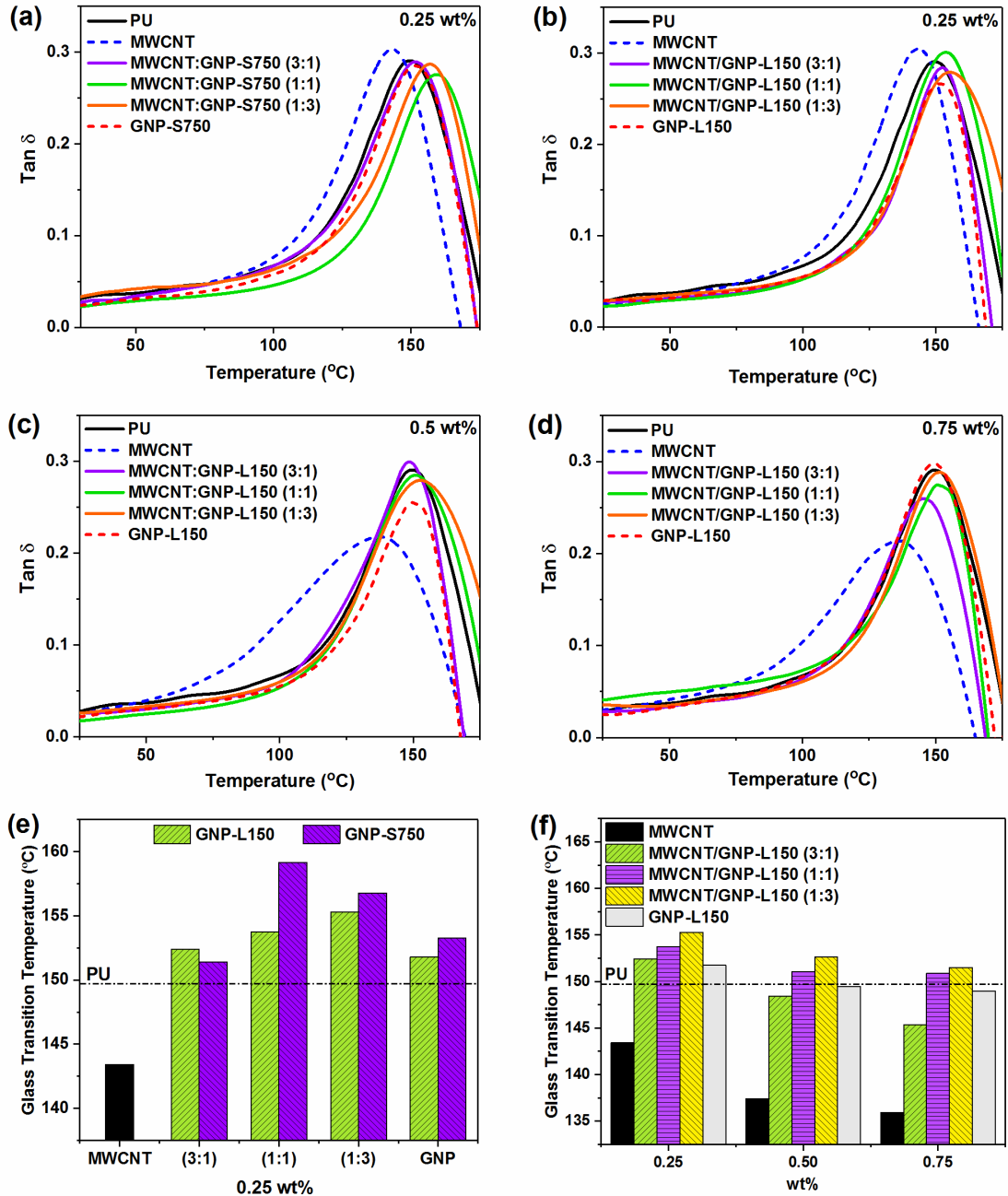

Fig. 7. Tan δ of (a-d) hybrid MWCNTs/graphene nanocomposites versus temperature and (e-f) glass transition temperature of nanocomposites.

### 3.3.3 C-Factor

The carbon nanofillers effectiveness on PU bases nanocomposites can be described using the C-factor coefficient (C) using the storage modulus of PU and its nanocomposites, which offers a closer outline about the interaction of nanofillers and the matrix, given by [38, 48]:

$$C = \frac{\left[E'_g / E'_r\right]_{Nanocomposites}}{\left[E'_g / E'_r\right]_{PU}} \quad (1)$$

where $E'_r$ and $E'_g$ denote the storage modulus in the rubbery and the glassy regions, respectively. The C-factor value indicates a ratio of storage modulus in the glassy to rubbery regions, which is the inverse of nanofillers effectiveness in the polymer matrix. Therefore, the lower value of the C-factor represents higher nanofillers efficacy along with their better distribution within the PU matrix. The C-factor value for hybrid and single nanocomposites are illustrated in Fig. 8(a). Hybrid MWCNTs/GNP-S750 based nanocomposite has the lowest C-factor value, which revealed the maximum nanofillers effectiveness, in comparison with solely MWCNTs and graphene.

### 3.3.4 Degree of Entanglement

The degree of entanglement in the nanocomposites can be measured using DMA, which depends on the interaction of the nanofillers and the polymer matrix. The maximum nanofillers effectiveness in the nanocomposites was evaluated via the degree of entanglement using the storage modulus value, as follow [38, 49]:

$$N = \frac{E'}{6RT} \quad (2)$$

where $E'$, $T$ and $R$ are the storage modulus at the rubbery region (175 °C), the absolute temperature and the universal gas constant, respectively. Fig. 8(b) illustrates the degree of entanglement between single or hybrid carbon nanofillers and PU matrix. Hybrid nanocomposites represented higher $N$ values, where the maximum entanglement density (31.17 mol/m$^3$) was achieved in PU/MWCNT+GNP-S750 due to the physical adhesion between MWCNTs and GNP-S750, as well as the better interfacial interaction among hybrid nanofillers and PU matrix [48]. It can be concluded that mechanical loading efficiently transferred from the PU matrix to 3D hybrid MWCNTs/GNP-S750 as a result of dominant interfacial interaction.

### 3.3.5 Cross-Link Density

The contribution of nanofillers and the PU matrix to the storage modulus of the nanocomposites that are arising from the nanofiller-matrix interface was calculated using the Maier and Goritz model, as follow [50]:

$$E' = \lambda K_B T \tag{3}$$

where $\lambda$, $K_B$ and $T$ represent cross-link density, the Boltzmann constant and the temperature in the Kelvin scale (25 °C), respectively. The cross-link density of single and hybrid nanocomposites at 0.25 wt% is illustrated in Fig. 8(c). The cross-link density of neat PU is $2 \times 10^{27}$ m$^{-3}$, whereas the MWCNTs/GNP-S750 based hybrid nanocomposite presented the cross-link density of $3.72 \times 10^{27}$ m$^{-3}$, which is higher than the value of other nanocomposites due to the synergistic effect between GNP-S750 and MWCNTs. It can be concluded that effective dispersion of MWCNTs/GNP-S750 formed highly cross-linked nanofillers within the PU matrix [48]. Moreover, the interconnections of hybrid nanofillers developed further crosslinks, which enhanced the elastic properties [50].

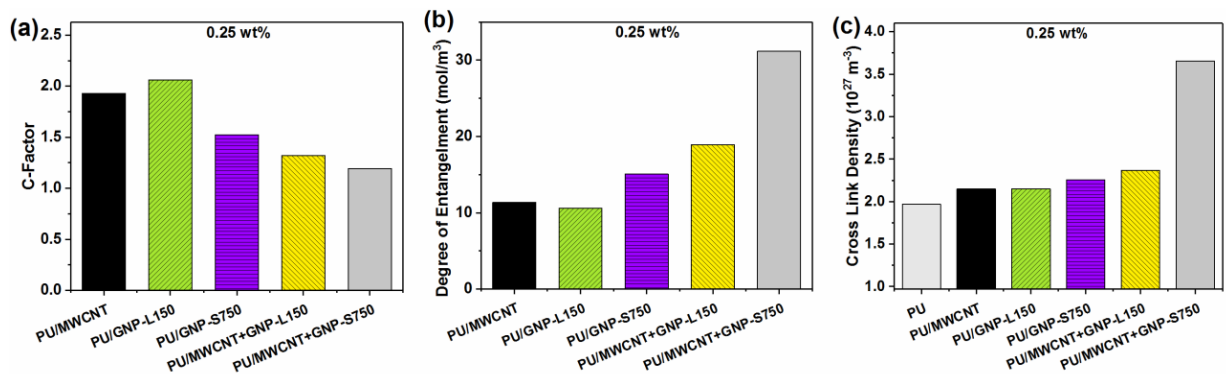

Fig. 8. (a) C-factor, (b) the degree of entanglement and (c) cross-link density of the nanocomposites.

## 4. Conclusion

PU foam nanocomposites including MWCNTs and two variety of graphene were fabricated at low nanofillers loadings and various ratios via a simple, quick and scalable method. TGA, SEM, TEM, Raman spectroscopy and DMA have been performed to evaluate the most substantial enhancement in thermomechanical properties of PU foams. Uniform distribution of the interconnected graphene through CNTs was confirmed using SEM and TEM images. Accordingly, PU foams including solely GNP-S750 or both graphene and MWCNTs had much higher thermal stability relative to pure PU and the samples containing only MWCNTs. The viscoelastic properties of PU nanocomposites were carried out using DMA measurement, in which storage modulus, damping factor (Tan δ) and the glass transition temperature ($T_g$) were studied. The storage modulus of PU improved with single nanofillers, being higher for PU/GNP-S750 at 0.75 wt% contents with 22% enhancement, whereas the $T_g$ value diminished at higher concentrations. Results showed that the combination of MWCNTs and graphene in the PU matrix exhibited a synergistic effect regarding storage modulus and $T_g$, which exceeded the influence of each nanofiller. In comparison with pure PU, the storage modulus of hybrid MWCNTs/GNP-S750 (1:1) nanocomposite improved about 86% at a low content of 0.25 wt%, while PU/MWCNTs and PU/GNP-S750 showed a 15% and 19% enhancement at the same loading, respectively. Moreover, the highest storage modulus of PU nanocomposites using single MWCNTs or graphene can be found at 0.75 wt% for GNP-S750 with a 22% improvement. A 3D hybrid network was formed in hybrid nanocomposites, in which MWCNTs interconnected adjacent graphene. When the graphene /MWCNTs hybrid nanocomposites were stretched, the nanofillers distance enhanced, however, the initial nanofillers condition was found again after relaxation, which led to enhancing the storage modulus of PU. The $T_g$ value of PU was also synergically enhanced by about 9.5 °C with 0.25 wt% inclusion of well-dispersed 3D MWCNT/GNP-S750 (1:1). The C-factor (reinforcement coefficient), degree of entanglement and cross-link density of fabricated nanocomposites were also examined using the storage modulus value to evaluate the interaction of single and hybrid carbon nanofillers with PU matrix. All of these outcomes revealed that the GNP-S750 with a smaller dimension, a higher SSA and more defects have a higher ability to form a more effective 3D architecture, which further improved the storage modulus and $T_g$ of PU when combined with MWCNTs as hybrid nanofillers. Here, these hybrid graphene /CNTs nanocomposites are favorable for lightweight foam applications that thermal stability and thermomechanical properties are crucial principles, such as packaging and structural damping materials.

## CRediT authorship contribution statement

**Amir Navidfar:** Conceptualization, Methodology, Validation, Formal analysis, Investigation, Writing - Original draft, Visualization. **Tugba Baytak:** Performing DMA tests. **Osman Bulut:** Performing DMA tests, Validation, Review & Editing. **Hikmet Iskender:** Validation, Review & Editing. **Levent Trabzon:** Conceptualization, Methodology, Validation, Resources, Review & Editing, Supervision, Project administration.